 \documentclass[preprint2]{aastex}

\newcommand{\msun}{{M$_\sun$}}

\newcommand{\ergs}{{erg s$^{-1}$}}

\newcommand{\x}{{\rm{X}}}

\newcommand{\nuno}{NGC~4485}
\newcommand{\ndos}{NGC~4490}
\newcommand{\nboth}{NGC~4485/90}
\newcommand{\m}{$\mu$m}

\shorttitle{IRS spectroscopy of ULXs in NGC~4485/90}
\shortauthors{V\'azquez G. A. et al.}

\begin{document}


\title{Constraints on accretion in Ultraluminous X-ray Sources \\
from {\em Spitzer} IRS observations of NGC~4485/90: Infrared diagnostic diagrams}


\author{Gerardo A. V\'azquez}
\affil{Physics \& Astronomy Department, Johns Hopkins University,
    Baltimore, MD 21218}
\email{vazquez@pha.jhu.edu}

\author{Ann E. Hornschemeier}
\affil{NASA/GSFC, Code 662, Laboratory for X-ray Astrophysics
Greenbelt, MD 20771}
\email{annh@milkyway.gsfc.nasa.gov}

\author{Edward Colbert}
\affil{Physics \& Astronomy Department, Johns Hopkins University,
    Baltimore, MD 21218}
\email{colbert@jhu.edu}

\author{Timothy P. Roberts and Martin J. Ward}
\affil{Dept. of Physics, Durham University, South Road, Durham DH1, 3LE, UK}
\email{t.p.roberts@durham.ac.uk, m.ward@durham.ac.uk}


\author{Sangeeta Malhotra}
\affil{Arizona State University, Department of Physics \& Astronomy, P.O. Box 871504
Tempe, AZ 85287-1504}
\email{sangeeta.malhotra@asu.edu}




\begin{abstract}

Constraining the astrophysical nature of Ultra-Luminous X-ray (ULX) sources, which have X-ray luminosities exceeding 10$^{39}$~\ergs, has been elusive due to the optical faintness of any counterparts. With high spectral resolution observations in the $\sim10$--30$\mu$m wavelength range we have conducted an experiment to study six ULX sources in the NGC~4485/90 galaxy pair.  We have found that five of the six ULXs, based on mid-infrared spectral diagnostics, show the characteristic higher ionization features that are found in AGN.  The sixth source, ULX-1, is consistent with being a supernova remnant.  The chief infrared spectral diagnostics used are the ratios of [S~{\small III}]/[Si~{\small II}] {\em vs} [Ne~{\small III}]/[Ne~{\small II}]. In two instances fits to the continuum and {\em poly aromatic hydrocarbons} (PAH) features also indicate higher dust temperatures that are characteristic of accreting sources.  Overall, however, we find the continuum is dominated by stellar processes, and the best diagnostic features are the emission lines.  High spectral resolution studies in the mid-infrared thus appear to show great promise for determining the astrophysical nature of ULXs.

\end{abstract}


\keywords{galaxies: general --- ULRGS: individual(NGC 4485,
NGC 4490)}



\section{Introduction}

Ultra-Luminous X-ray sources (ULXs) are generally defined as extra-nuclear  galactic X-ray sources with X-ray luminosities $L_{\x} \geq 10^{39}$~{\ergs}, a luminosity that is seldom reached by any Galactic (Milky Way) X-ray binary (XRB). If ULXs emit their X-rays isotropically, by Eddington arguments, the central object could be a black hole (BH) with mass M $\ge$ 5~{\msun}, and for Eddington ratios of $\sim$10$^{-2}$$-$10$^{-1}$, the implied mass of the central object is $\sim$50$-$500~{\msun}.  Therefore, some ULXs could be accreting intermediate-mass black holes \citep[IMBHs,][]{colmus99,ptagri99,millcol04}. The majority of ULXs, however, are likely accreting {\it stellar} mass black holes (e.g., $\sim1$--10~{\msun}) as there appears to be a strong association between ULX sources and star formation in galaxies \citep{zesfab02}. One popular model claims that ULXs are merely stellar-mass BHs in high-mass XRBs, and mild beaming of their X-ray luminosities produces the observational illusion of extremely high X-ray luminosities \citep[e.g.,][]{king01,king06}. Hundreds of ULXs are known at the present time (e.g., Roberts \& Warwick 2000, Colbert \& Ptak 2002, and Chandra ULX catalogs by Ptak et al.\footnote{URL: http://www.xassist.org}, Swartz et al. 2004, and Liu \& Bregman 2005). 

Accreting binaries such as these could be precursors to eventual gravitational radiation events. Further, ULXs are important as they often dominate the total observed X-ray flux in starburst galaxies \citep[see][]{colbert04}.  Their relationship with star formation also appears to extend to cosmologically interesting ($z \gtrsim 0.1$) look-back times \citep{horns04,lehmer06} and may provide a cosmological tool. 

Unfortunately, all but the very brightest ULX sources are too faint for detailed X-ray spectroscopy studies \citep[the best cases must rely upon resolving power of R$=60$ at 6.4~kev, available with X-ray CCDs;][]{mizefa04}. 

There is perhaps hope for ascertaining the nature of ULX sources by studying their interactions with their surroundings, such as gas, dust and stars; such information may be captured using infrared spectral diagnostics.   \citet{genzel98} were the first to show that ionization-sensitive indices based on mid-infrared ratios were helpful for identifying
the nature of heavily obscured nuclear sources \citep{genzel98,laurent00,sturm02,peeters04}. We have thus taken advantage of the resolving power (R~$\sim600$) of the {\em Infrared Spectrograph} (IRS) on board the {\em Spitzer Space Telescope} (SST) to calculate highly accurate ionization diagnostic ratios.

The nearby galaxy pair NGC~4485/90 provides an excellent opportunity for studying ULX sources \citep[see][]{roberts02}. There are six ULXs in NGC~4485/90 \citep[D $=$ 7.8~Mpc,][]{tully88}, far more than is typical \citep[$\le$ 2 in most cases, and 0.2, on average;][]{colpta02,ptacol04}. 
The unprecedented sensitivity and angular resolution afforded by the {\em Spitzer} Space Telescope's Infrared Spectrograph (IRS) coupled with the proximity of NGC~4485/90 allow us to carry out this important experiment.

\section{IRS spectroscopy: Observing strategy and Reduction}

We have performed spectral mapping of the system {\nboth} by using the mapping mode of the IRS \citep{houck04} on board the SST. Rectangular regions of $11\farcs1{\times}22\farcs3$ and $4\farcs7{\times}11\farcs3$ (i.e. zones of $\sim 0.42 \times 0.84$~kpc and $\sim 0.18 \times 0.42$~kpc at the system) were mapped using both the short-high (SH, $\sim 9.9 - 19.7$~{\m}) and long-high (LH, $\sim 19.5-38$~{\m}) slits, respectively. Six rectangular regions centered at the positions of the ULXs (Table~\ref{tbl-1}) were covered using a $1{\times}4$ and $1{\times}7$ grid array for the long and and short slits of the IRS. In addition two positions in {\ndos} that are {\em not} near ULXs were observed with the same configuration as the one used for the ULXs. The integration time was 240~s at each position, and the step size is half the slit width perpendicular to the slit. This configuration allows net exposures of ${\ge 480}$~s for rectangular sizes $16.7{\times}22.3$'' and $14.1{\times}11.3$'' for long and short high slits, respectively. 

Sets of 8 exposures were taken for each position with the IRS. The pre-reduced basic calibrated data ({\em bcd}) files were obtained from the {\it Spitzer Science Center} (SSC) data base. In order to extract the spectra for the SH and LH modes, we used SMART \citep[Spectroscopy Modeling Analysis and Reduction,][]{higdon04} which is publicly available software provided by the SSC\footnote{http://ssc.spitzer.caltech.edu/archanaly/contributed/smart/}. 

\section{Infrared Spectral Diagnostics}

\begin{figure}
\plotone{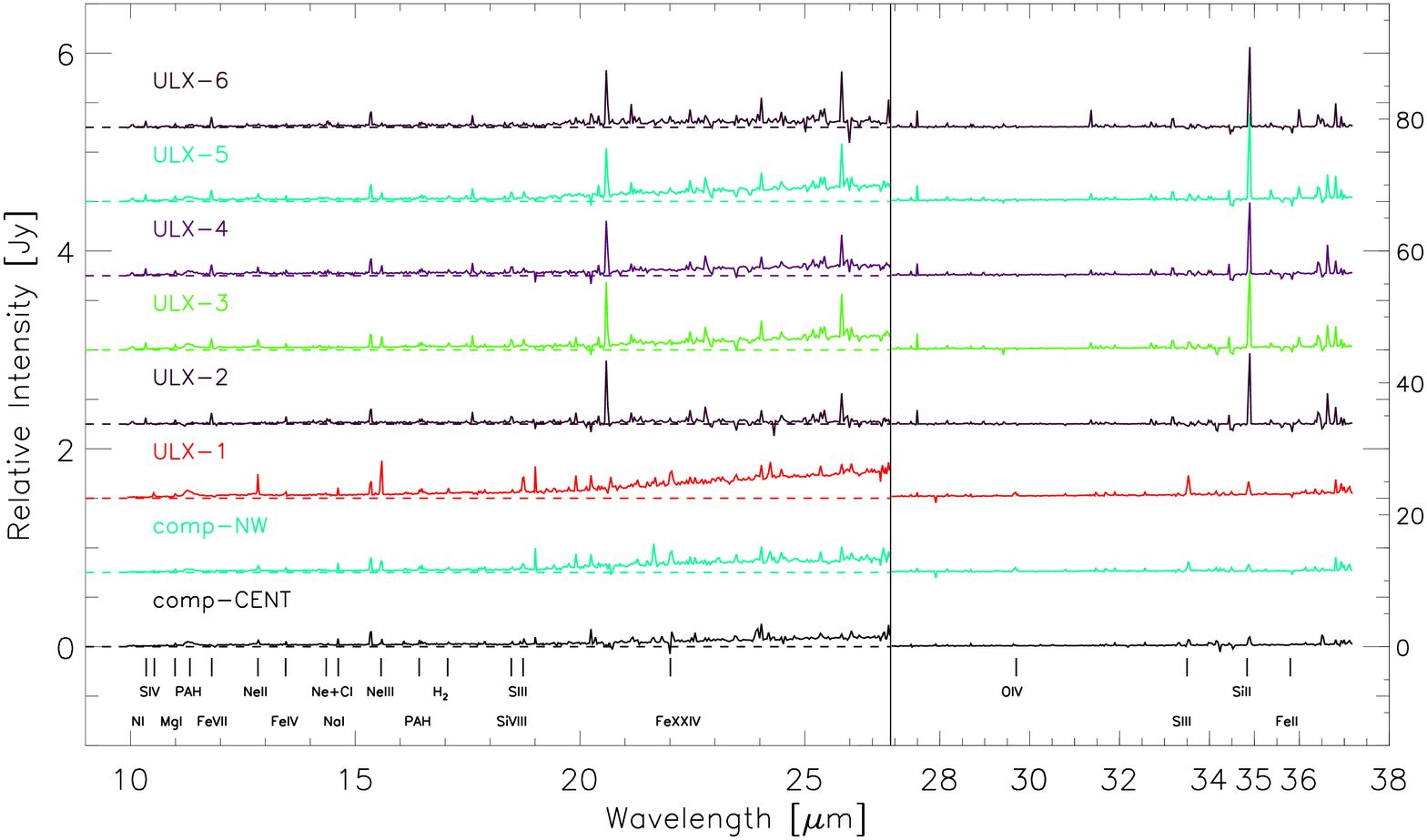}
\caption{The combined spectra with the left SH and right LH modes for all of the sources analyzed in this study. For better visualization, the spectra have been plotted displaced along the y-axis. From $\sim9$ to $\sim 26.9$ {\m} spectra are displaced by 0.75 Jy and $\sim26.9$ to $\sim38$ {\m} spectra are displaced by 15 Jy. Dashed lines are set to help indicate the slope of the continuum at wavelengths $\ge 20$~{\m}.
\label{fig1}}
\end{figure}

We have used line identifications from \citet{smith04}, \citet{armus06} and NIST\footnote{Available at http://physics.nist.gov/cgi-bin/AtData/main{\_}asd/} \citep{mar95}. We have combined both modes by re-scaling the LH mode spectra to the SH mode spectra by adding or subtracting a constant and then re-binning to a resolution of 0.0287~$\mu$m/pixel. In Fig.~\ref{fig1} we show the combined spectra from all the positions. 

As we can see from Fig.~\ref{fig1}, some of the line features are present or stronger in the ULX region than in the comparison region spectra. The most striking examples are features such as [S~{\small IV}]10.51, [Ne~{\small II}]12.81, [Na~{\small I}]14.6, [Ne~{\small III}]15.55, [S~{\small III}]18.71, and [S~{\small III}]33.48~{\m} that seem to be significantly stronger in the comparison regions and ULX-1(which is thought to be a SNR rather than an accreting binary), than in the rest of the sources. In the opposite way, the [Si~{\small II}]34.82~{\m} line is stronger in ULX-2 to ULX-6.

The purpose of this work is to generate some emission line diagnostic diagrams similar to commonly used optical emission line diagrams, which have proven useful in separating accretion-dominated sources from those dominated by stellar sources \citep[e.g.,][and references therein]{baldwin81,kauff06}. A popular mid-infrared diagnostic first put forth in an ISO spectroscopic survey of ultra-luminous infrared galaxies (ULIRGs) by \citet{genzel98}, and later explored by \citet{peeters04}, plots the emission line ratio [O~{\small IV}]25.91 {\m} $/$ [Ne~{\small II}]12.81 {\m} versus the strength of a mid-infrared {\em poly aromatic hydrocarbons} (PAH) features. 
These two lines however, are relatively weak and difficult to detect in lower luminosity systems. Taking advantage of the strong low ionization line emission of [Si~{\small II}]34.82 {\m}, \citet{dale06} suggested alternative diagrams to distinguish AGNs from SF regions such as [Ne~{\small III}]15.55{\m}$/$[Ne~{\small II}]12.81{\m} and [S~{\small III}]33.48{\m}$/$[Si~{\small II}]34.82{\m}. These ratios proved to be successful in separating accretion-powered vs. SF powered regions in the SINGS galaxy sample.

We have calculated these ratios for our eight spectra, shown in Fig.~\ref{fig2}. We have also plotted the data from \citet{dale06} that include nuclear and extra-nuclear regions from SINGS, the SMC/LMC and galactic H~{\small II} regions. Errors in the emission line fluxes calculation are around 10{\%}.
The boundaries in Fig.~\ref{fig2} are defined with the same curves as in \citet{dale06}. As we can see from Fig.~\ref{fig2}, ULX-2, ULX-3, ULX-4, ULX-5, and ULX-6 are in zone I where Seyfert and Liners are observed, while ULX-1, comp-NW, and comp-CENT are located in zone III as star forming systems. For ULX-2 and ULX-6 the line ratio [S {\small III}]33.48~{\m}$/$[Si~{\small II}]34.82~{\m} is $ < 0.01$, which is extremely low compared with extragalactic objects in Fig.~\ref{fig2}. 


The SINGS group has recently developed models \citep[{\em pahfit}, ][]{smithjd07} to characterize PAH features. The models include starlight continuum, featureless thermal dust continuum, pure rotational lines of H$_2$, fine-structure lines, dust emission features, and dust extinction. The starlight continuum is represented by a blackbody emission was fixed at $T_\star = 5000$ K. Nine thermal dust continuum components are represented by blackbodies at fixed temperatures $T_{\rm m} = 35, 50, 75, 90, 110, 135, 200, 300, 500$ K. The program {\em pahfit} performs a fit through two hundred iterations, each iteration determines a $\chi^2$ value for all nine temperatures, until a minimum value for $\chi^2$ is found. Dust temperatures are based on the shape of the continuum spectrum. We show the corresponding fits in Table~\ref{tbl-1}, where we show ULX-2 and ULX-6 have hotter components, but the relative intensity of those components is low for ULX-1.
 
The presence of a hotter dust component will flatten the spectra for our range of wavelengths. This effect is seen in Fig~\ref{fig1} for ULX-2 and ULX-6.The rest of the parameters modeled will be subject of a forthcoming paper (V\'azquez, et al. 2007b, {\em in preparation}).

Finally, we have found a strong linear anti-correlation between the observed $\log(L_{\rm X})$ and $\log$([S~{\small III}]33.48~{\m}$/$[Si~{\small II}]34.82~{\m}). The regression and intercept coefficients are $-0.21 \pm 0.05$ and $39.09 \pm 0.6$, respectively with a RMS = 0.08.

\section{Discussion}

\begin{figure}
\plotone{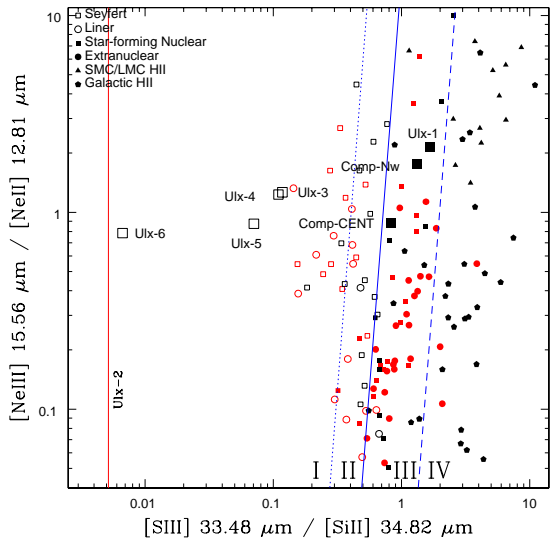}
\caption{A neon, sulfur, and silicon diagnostic diagram showing line ratios for the eight regions in the NGC~4485/90 system (names in black). Nuclear and extra-nuclear regions from the sample of SINGS \citep{dale06} are also plotted (in red) together with SMC, LMC and galactic H~{\small II} regions. The red line for ULX-2 means that the value for the line ratio of [Ne~{\small III}]15.55~{\m}$/$[Ne~{\small II}]12.81~{\m} was not determined due to  the undetected lines of Ne for this source. Zones III and IV are typical of SF-regions. Objects in zones I and II, can be classified as AGN-powered \citep{dale06}.
\label{fig2}}
\end{figure}

For the first time the IR continuum and lines ($\sim9.9-38$~{\m}) have been observed in ULXs by using the power of the {\em Spitzer} IRS. It is likely that this ULXs are irradiating nearby star-forming clouds. Although the geometry is probably thus quite complex, we do find some hints of information on the underlaying X-ray radiation. The most striking result from the data analyzed in this work is the one in Fig.~\ref{fig2}, which shows that five of the six ULXs reported by \citet{roberts02} fall in the ``AGN zone'' determined by the infrared diagnostic emission line ratios defined by the emission lines of  Ne, S, and Si \citep{dale06}. Both of the comparison regions used here, which do not harbor ULX sources, fall in the zone of SF. The SF zone is also occupied by extra-nuclear star-forming regions in the SINGS galaxy sample and SMC/LMC  and galactic H~{\small II} regions. The emission line at [N~{\small I}]10.33 and other unidentified features also behave like [Si~{\small II}]34.82~{\m}, showing a stronger contribution in ULX-2 to 6 than in ULX-1 and both of the comparison regions. 

The presence of a hotter dust component in our PAH/continuum modeling (see Table~\ref{tbl-1}), suggests that at least for ULX-2 and ULX-6 the [Si~{\small II}] lines could be originating from an accreting source. A dilution factor could be affecting the rest of the ULXs possibly due to their location within their host galaxy. All of this is consistent with the anti-correlation we have found between the line emission ratio of [S~{\small III}]33.48{\m}$/$[Si~{\small II}]34.82{\m} and the observed X-ray luminosity of the six ULXs, the more luminous the sources the lower the ratio of these emission lines. 

Different scenarios for the strong [Si~{\small II}] line have been suggested in the literature. In these scenarios the increasing density in H~{\small II}, photo-dissociation (PDR), and X-ray dissociation regions (XDR), respectively, increases the intensity of the [Si~{\small II}]34.82~{\m} line \citep[see][]{maloney96,kaufman06,meij05}. A third scenario suggests that heavy elements such as Si, Mg, and Fe may be returned to the gas phase by dust destruction (e.g., sputtering) in regions subject to strong shocks caused by stellar winds, starbursts, and AGN activity \citep{dale06}. Details of the geometry and ionization balance in the region require more detailed fitting, and also spatial mapping of spectral parameters as ULXs are thought to have drifted from their birthplaces. The origin of these lines could thus be disentangled with detailed IRS mapping observations and IRAC imaging, which are the subject of a forthcoming paper (V\'azquez et al. 2007b).


With the exception of a weak detection in ULX-1, we have not detected the emission line [O~{\small IV}]25.91~{\m}, this line is a strong feature in SNR \citep[e.g.,][]{morris06,williams06}, supporting previous suggestions that ULX-1 is a SNR \citep[e.g., based on radio and X-ray constraints;][]{roberts02}. 

The mid-infrared emission line diagnostics used to uncover the nature of AGNs and SF galaxies has been shown here to work well separating SFRs and young SN from probably true accreting ULXs \citep[according to X-ray observations;][]{roberts02}. But this is a small sample of ULXs to claim a general rule that could be improved observing a higher number of ULXs. The present work shows that a new diagnostic window has been opened on ULXs and their immediate environment via important mid-infrared features like [Si~{\small II}]34.82 {\m}, made observationally available by the extraordinary performance of the IRS on board the {\em Spitzer} Space Telescope.

\acknowledgments

Many thanks to Danny A. Dale who provided us his data for our figures. This work has been supported by the {\em Spitzer} grant 3360 (P.I. Colbert).

\clearpage

\clearpage

\clearpage

\begin{deluxetable}{cccccccc}
\tabletypesize{\scriptsize}
\tablecaption{The list of ULXs of the {\nboth} system taken from \citet{roberts02}.
\label{tbl-1}}
\tablewidth{0pt}
\tablehead{
\colhead{ULX} & \colhead{RA (J2000)} & \colhead{DEC (J2000)} & \colhead{Notes} & \colhead{$L_{\rm X}$\tablenotemark{a}} & \colhead{$T_{dust}$} & \colhead{$\tau$\tablenotemark{b}} & \colhead{$\tau_{9.7}$} \\
\colhead{} & \colhead{} & \colhead{} & \colhead{} & \colhead{$10^{39}$~{\ergs}} & \colhead{[K]} & \colhead{[$10^{-10}$]} & \colhead{[$10^{-3}$]}
}

\startdata
1  &  12:30:29.5  &  +41:39:27  &  {\ndos} Radio Src. Possible SNR & 1.0 (4.9) & 50, 75, & 72000, 13000  & 3310 \\
   &    &   &   &   & 200, 300 & 0.1, 1.1 & \\
2  &  12:30:30.6  &  +41:41:42  &  {\nuno} X-1 & 4.0 (4.6) & 135, 200 & 56, 2.3 & 4810\\
3  &  12:30:30.8  &  +41:39:11  &  {\ndos} Transient X-ray Src. & 2.9 (4.6) & 90, 110 & 740, 240 & 3.93 \\
4  &  12:30:32.3  &  +41:39:18  &  {\ndos} X-1 & 1.9 (2.6) & 75, 90 & 2500, 960 & 1.75\\
5  &  12:30:36.3  &  +41:38:37  &  {\ndos} X-2 & 1.9 (2.6) & 75, 90 & 8800, 16 & 7.94 \\
6  &  12:30:43.2  &  +41:38:18  &  {\ndos} X-4 & 3.1 (4.4) & 75, 90 & 360, 580 & 2990 \\
   &   &   &   &   & 110, 135 & 190, 10 & \\

\enddata
\tablenotetext{a}{Numbers in parenthesis give the intrinsic (unabsorbed) luminosity.}
\tablenotetext{b}{The total extinction as a function of the extinction at 9.7 {\m} ($\tau_{9.7}$). The total extinction is modeled as a power low plus silicate features at 9.7 and 18~{\m}. The features are counted as a combination of measured galactic extinction profiles for these wavelengths \citep{smithjd07}.}
\end{deluxetable}


\begin{thebibliography}{}
\bibitem[Armus et al.(2006)]{armus06} Armus, L., et al. 2006, \apj, 640, 204
\bibitem[Baldwin, Phillips \& Terlevich (1981)]{baldwin81} Baldwin, J. A., Phillips, M. M., \& Terlevich, R. 1981, \pasp, 93, 5
\bibitem[Colbert et al.(2004)]{colbert04} Colbert, E., et al.  2004, \apj, 602, 231
\bibitem[Colbert \& Mushotzky(1999)]{colmus99} Colbert, E., \& Mushotzky, R. 1999, \apj, 519, 89 
\bibitem[Colbert \& Ptak(2002)]{colpta02} Colbert, E.,  \& Ptak, A. 2002, \apjs, 143, 25
\bibitem[Dale et al.(2006)]{dale06} Dale, D. A. et al. 2006, \apj, 646, 161
\bibitem[Genzel et al.(1998)]{genzel98} Genzel, R. et al. 1998, \apj, 498, 579
\bibitem[Higdon et al.(2004)]{higdon04} Higdon, S. J. U., et al. 2004, \pasp, 116, 975
\bibitem[Hornschemeier et al.(2004)]{horns04} Hornschemeier, A. E., et al. 2004, \apjl, 600, 147
\bibitem[Houck et al.(2004)]{houck04} Houck, J. R., et al. 2004, \apjs, 154, 18
\bibitem[Kaufman et al.(2006)]{kaufman06} Kaufman, M. J., Wolfire, M. G., \& Hollenbach, D. J. 2006, \apj, 644, 283
\bibitem[Kauffmann et al.(2006)]{kauff06} Kauffmann, G., Heckman, T. M., De Lucia, G., Brinchmann, J., Charlot, S., Tremonti, C., White, S. D. M., \& Brinkmann, J. 2006, \mnras, 367, 1394
\bibitem[King et al.(2001)]{king01} King, A. R., et al. 2001, \apj, 552, 109
\bibitem[King (2006)]{king06} King, A. R. 2006, in Compact stellar X-ray sources. Edited by Walter Lewin \& Michiel van der Klis. Cambridge Astrophysics Series, No. 39. Cambridge, UK: Cambridge University Press, p. 507
\bibitem[Laurent et al.(2000)]{laurent00} Laurent, O., et al. 2000, \aap, 359, 887
\bibitem[Lehmer et al.(2006)]{lehmer06} Lehmer, B. D., Brandt, W. N., Hornschemeier, A. E., Alexander, D. M., Bauer, F. E., Koekemoer, A. M., Schneider, D. P., \& Steffen, A. T. 2006. \aj, 131, 2394 
\bibitem[Liu \& Bregman(2005)]{liubreg05} Liu, Ji-Feng, Bregman, J. N. 2005, \apjs, 157, 59
\bibitem[Maloney, Hollenbach, \& Tielens(1996)]{maloney96} Maloney, P. R., Hollenbach, D. J., \& Tielens, A. G. G. M. 1996, \apj, 466, 561
\bibitem[Martin et al.(1995)]{mar95} Martin, W. C., Sugar, J., Musgrove, A., 
Dalton, G. R., Wiese, W. L., \& Fuhr, J. R. 1995, NIST Database for Atomic 
Spectroscopy, Version 2.0, NIST Standard Reference Database 61 
(Gaithersburg: NIST)
\bibitem[Meijerink \& Spaans(2005)]{meij05} Meijerink, R., \& Spaans, M. 2005, \aap, 436, 397
\bibitem[Miller \& Colbert(2004)]{millcol04} Miller, C. M., \& Colbert, E. J. M. 2004, Int. J. Mod. Phys. D, Vol. 13, No. 1, p. 1 (astro-ph/0308402)
\bibitem[Miller et al.(2004)]{mizefa04} Miller, J. M., Zezas, A., Fabbiano, G., \& Schweizer, F. 2004, \apj, 609, 728
\bibitem[Morris et al.(2006)]{morris06} Morris, P. W., et al. 2006, \apj, 640, L179
\bibitem[Peeters, Spoon, \& Tielens(2004b)]{peeters04} Peeters, E., Spoon, H. W. W., \& Tielens, A. G. G. M. 2004b, \apj, 613, 986
\bibitem[Ptak \& Colbert(2004)]{ptacol04} Ptak, A. F., \& Colbert, E. J. M. 2004, \apj, 606, 291
\bibitem[Ptak \& Griffiths(1999)]{ptagri99} Ptak, A., \& Griffiths, R. 1999, \apj, 517, 85
\bibitem[Roberts et al.(2002)]{roberts02} Roberts, T., et al. 2002, \mnras, 337, 677
\bibitem[Roberts \& Warwick(2000)]{robwar00} Roberts, T., \& Warwick, R. 2000, \mnras, 315, 98
\bibitem[Smith et al.(2004)]{smith04} Smith, J. D. T., et al. 2004 \apjs, 154, 159
\bibitem[Smith et al.(2007)]{smithjd07} Smith, J. D. T., et al. 2007 \apj, accepted
\bibitem[Sturm et al.(2002)]{sturm02} Sturm, E., et al. 2002, \aap, 393, 821
\bibitem[Swartz et al.(2004)]{swartz04} Swartz, D. A., Ghosh, K. K., Tennant, A. F., \& Wu, K. 2004, \apj, 154, 519
\bibitem[Tully(1998)]{tully88} Tully, R. B. 1988, Catalogue of Nearby Galaxies, Cambridge and New York, Cambridge University Press, 1988, p 221
\bibitem[Williams, Chu, \& Gruendl(2006)]{williams06} Williams, R. M., Chu, Y. H., Gruendl, R. 2006, \aj, 132, 1877
\bibitem[Zezas \& Fabbiano(2002)]{zesfab02} Zezas, A., \& Fabbiano, G. 2002b, \apj, 577, 726 

\end{thebibliography}
\end{document}